\documentclass[aps,pra,showpacs,twocolumn,amsmath,amssymb,superscriptaddress,footinbib]{revtex4}

\usepackage[english]{babel}
\usepackage{latexsym}
\usepackage{graphics,psfrag}
\usepackage{epsfig}

\def\be{\begin{equation}}
\def\ee{\end{equation}}
\def\bea{\begin{eqnarray}}
\def\eea{\end{eqnarray}}
\def\bi{\begin{itemize}}
\def\ei{\end{itemize}}
\def\bin{\begin{enumerate}}
\def\ein{\end{enumerate}}
\def\la{\langle}
\def\ra{\rangle}

\newcommand{\hl}{\xi}

\begin{document}
\title{Quantum bright soliton in a disorder potential}

\author{Krzysztof Sacha}
\affiliation{
Instytut Fizyki imienia Mariana Smoluchowskiego and
Mark Kac Complex Systems Research Center, 
Uniwersytet Jagiello\'nski, ulica Reymonta 4, PL-30-059 Krak\'ow, Poland}

\affiliation{Laboratoire Kastler-Brossel, UPMC, ENS, CNRS;  
4 Place Jussieu, F-75005 Paris, France}

\author{Dominique Delande}

\affiliation{Laboratoire Kastler-Brossel, UPMC, ENS, CNRS;  
4 Place Jussieu, F-75005 Paris, France}
 
\author{Jakub Zakrzewski} 

\affiliation{
Instytut Fizyki imienia Mariana Smoluchowskiego and
Mark Kac Complex Systems Research Center, 
Uniwersytet Jagiello\'nski, ulica Reymonta 4, PL-30-059 Krak\'ow, Poland}

\affiliation{Laboratoire Kastler-Brossel, UPMC, ENS, CNRS;  
4 Place Jussieu, F-75005 Paris, France}

\date{\today}

\begin{abstract}At very low temperature, a quasi-one-dimensional ensemble of atoms with attractive
interactions tend to form a bright soliton.
When exposed to a sufficiently weak external potential, the shape
of the soliton is not modified, but its external motion is affected. We develop in detail
the Bogoliubov approach for the problem, treating, in a non-perturbative way, the motion of the
center of mass of the soliton. Quantization of this motion allows us to discuss its long time
properties. In particular,
in the presence of a disordered potential, the quantum motion of the center
of mass of a bright soliton may exhibit Anderson localization, on a localization length
which may be much larger than the soliton size and could be observed experimentally. 

\end{abstract}


\maketitle
\section{Introduction}

Anderson localization is a localization effect predicted to take place for a
wave propagating in a disordered potential~\cite{anderson1958}. 
It is due to multiply scattered waves
from random defects and yields exponentially localized density profiles, 
resulting in a complete suppression of the usual diffusive transport associated 
with incoherent wave scattering~\cite{lee1985}. While in the three dimensional world,
one may observe a transition between extended and localized states,
in a one-dimensional (1D) world, Anderson localization is a typical feature of the 
motion in a disordered potential~\cite{vantiggelen1999}.

Cold atoms form a wonderful toolbox for controlling parameters of the system under study \cite{jaksch}. It comes out as no surprise that attempts have been made for a direct observation of
the Anderson localization in cold atoms settings. Already the first attempts
\cite{bodzio,inguscio,fort,aspect,wir,wir2} have revealed that the presence of atomic interactions
may deeply affect the physics of the problem and make the observation of the 
localization non trivial. Further theoretical studies \cite{LSP2007,Lugan2007,skipetrov2008} were followed by successful observations
of the phenomenon  made possible by
going to the regime of very weakly interacting particles~\cite{Billy2008}. While in that
work a random speckle potential was used, in another attempt~\cite{ingu2008} a quasi-periodic
version of the potential using superposition of laser beams was created resulting
in the observation of Aubry-Andr\'e~\cite{AA} localization for noninteracting atoms.
 
Anderson localization is a one-body phenomenon, and it is important to understand
how it is modified when interactions between particles -- in our case, cold atoms -- are
taken into account. In the absence of any external potential, at zero temperature,
1D particles interacting attractively tend to cluster together, forming a bright soliton.
Explicit solutions of the many-body problem can be found 
for a contact interaction~\cite{McGuire64}.
Altogether, a bright soliton appears as a composite particle, whose position is given
by the center of mass of the constituting atoms and a mass equal to the sum of the mass
of the atoms (see next section). 
Using external potentials, it has been experimentally shown how to put solitons
in motion~\cite{brightexp}.
The purpose of this contribution is to discuss
what happens to a bright soliton exposed to a weak and smooth disordered potential \cite{kivshar,others}.
Of course,  if that potential was sufficiently strong, it could probably destroy the soliton
altogether, break it into pieces etc. We are, however, interested in the other limit
when the external potential is sufficiently weak and smooth
 not to perturb the soliton shape. It is then quite reasonable to expect that, if this weak
 potential is of  random nature (disorder) that the soliton as a composite particle, 
 undergoes
multiple scattering, diffusive motion and eventually Anderson localization. In a recent
short contribution~\cite{ours} we have shown that this is indeed the case by considering the
effective quantum motion of the soliton. The present work brings a detailed derivation
of the effective Hamiltonian applied before, and shows examples of
the corresponding localized eigenstates. It provides thus a complementary material
to our previous work~\cite{ours}.

\section{Mean field description}
\label{mean}

\subsection{Equations of motion for a bright soliton in a disorder potential}
\label{meana}

Consider an ensemble of cold atoms (bosons) with attractive interactions at zero temperature.
We assume a strong harmonic transverse confinement so
a one-dimensional approximation can be 
used. In the mean field approach, a $c$-number function $\phi$ takes the place of the 
bosonic field operator $\hat\psi$.  
 $\phi$ is a solution of the Gross-Pitaevskii equation
\be
\label{GP}
i\partial_t\phi=-\frac12 \partial_z^2\phi-|\phi|^2\phi,  
\ee
where we have adopted the following natural units for energy, length and time, 
respectively 
\bea
E_0&=&4m\omega_\perp^2a^2, \\
l_0&=&\frac{\hbar}{2|a|m\omega_\perp}, \\ 
t_0&=&\frac{\hbar}{4a^2m\omega_\perp^2}. 
\eea
The transverse harmonic 
confinement frequency is denoted by $\omega_\perp$, $a$ is the atomic 
$s$-wave scattering length, and $m$ the mass of an atom. We normalize $\phi$ 
to the  total number of particles $N$. 
Eq.\eqref{GP} admits a stationary bright soliton solution
$e^{-i\mu t}\phi_0$ \cite{zakharov}, where
\be
\phi_0(z-q)=\sqrt{\frac{N}{2\hl}}\frac{e^{-i\theta}}{\cosh[(z-q)/\hl]}, 
\label{bs}
\ee
the chemical potential $\mu=-N^2/8$ and the soliton width $\hl=2/N$. This bright
solitonic solution minimizes the energy functional
\be
E=\int dz \left[\frac12 |\partial_z \phi|^2-\frac{1}{2} |\phi|^4 -\mu
|\phi|^2\right].  
\label{enfunc}
\ee
Observe that eq.~(\ref{bs}) allows for an arbitrary center-of-mass (CM) position $q$ 
and an arbitrary global phase $\theta$.

Suppose the soliton is placed in a weak and smooth disorder potential, $V(z)$, with
variance $V_0^2$ and correlation length $\sigma_0$. We will 
concentrate on the case
when $\sigma_0<\hl$ but the approach we present is general. 
Linearization of the
Gross-Pitaevskii equation allows us to describe the perturbation of the soliton 
due to the presence of a weak potential \cite{castin}. Indeed, the substitution 
\be
e^{-i\mu t}[\phi_0+\delta\phi],
\ee
into \eqref{GP} supplemented with the
potential $V(z)$ leads to the following inhomogeneous time-dependent Bogoliubov
equations
\be
i\partial_t \left(\begin{array}{c} 
\delta\phi 
\\
\delta\phi^*
\end{array}\right)={\cal L}\left(\begin{array}{c} 
\delta\phi 
\\
\delta\phi^*
\end{array}\right)+\left(\begin{array}{c} 
S 
\\
-S^*
\end{array}\right),
\label{tbe}
\ee
where 
\bea
{\cal L}=\left(\begin{array}{cc} 
-\frac{1}{2}\partial^2_z -2|\phi_0|^2-\mu 
& 
-\phi_0^2
\\
\phi_0^{*2}
&
\frac12\partial^2_z +2|\phi_0|^2+\mu
\end{array}\right),
\eea
and 
\be
S=V(z)\;\phi_0(z-q).
\ee 
In Eq.~\eqref{tbe} we have neglected terms of order higher than ${\cal
O}(\delta\phi,V)$.
Solution of \eqref{tbe} can be expanded in right
eigenvectors and corresponding adjoint modes of the non-hermitian operator 
${\cal L}$. However, this operator is not diagonalizable 
\cite{castin,lewenstein,dziarmaga04}. 
For all eigenvectors $(u_n,v_n)$ corresponding to non-zero eigenvalues 
$E_n$, the adjoint modes are left eigenvectors 
of the $\cal L$. That is no longer true for the zero-eigenvalue modes. There are two 
zero modes in our system
\be
\left(\begin{array}{c}u_\theta \\ v_\theta\end{array}\right)
=i\partial_\theta\left(\begin{array}{c}\phi_0\\ \phi_0^*\end{array}\right),
\quad
\left(\begin{array}{c}u_q \\ v_q\end{array}\right)
=i\partial_q\left(\begin{array}{c}\phi_0\\ \phi_0^*\end{array}\right),
\ee 
which are related to a small modification of the global phase of the solution 
\eqref{bs} and to a small shift of the CM, respectively \cite{dziarmaga04,ours}.
As both modifications 
cost no energy they appear as zero modes of the ${\cal L}$ operator. Indeed, 
it is consistent
with quadratic expansion of the energy functional,
\be
E={\rm const}+\frac12 \int dz\; (\delta\phi^*,-\delta\phi)\;{\cal L}\;
\left(\begin{array}{c}\delta\phi\\ \delta\phi^*\end{array}\right),
\ee
where we see that contributions to soliton perturbation from zero modes 
do not change $E$. 
The modes adjoint to the zero modes are
\be
\left(\begin{array}{c}u_\theta^{\rm ad} \\ v_\theta^{\rm ad}\end{array}\right)
=\partial_N\left(\begin{array}{c}\phi_0\\ \phi_0^*\end{array}\right),
\quad
\left(\begin{array}{c}u_q^{\rm ad} \\ v_q^{\rm ad}\end{array}\right)
=i\frac{z-q}{N}
\left(\begin{array}{c} 
\phi_0
\\
-\phi_0^*
\end{array}\right),
\label{adj}
\ee
which has been found by solving 
\be
{\cal L} \left(\begin{array}{c} u_{\theta,q}^\text{ad}\\ v_{\theta,q}^\text{ad}\end{array}\right)
=\frac{1}{M_{\theta,q}} 
\left(\begin{array}{c} u_{\theta,q}\\ v_{\theta,q}\end{array}\right),
\label{adjeq}
\ee
where $M_{\theta}$ and $M_q$ are determined by the requirements 
$\la u_{\theta}^\text{ad}| u_{\theta}\ra-\la v_{\theta}^\text{ad}| 
v_{\theta}\ra =1$ and 
$\la u_{q}^\text{ad}| u_{q}\ra-\la v_{q}^\text{ad}| 
v_{q}\ra =1$ \cite{castin,lewenstein,dziarmaga04,ours}. It turns out that 
\bea
M_\theta=-\frac{4}{N}, \quad M_q=N.
\label{masses}
\eea 
The latter is equal
to the total mass of the system.
Equation~\eqref{adjeq} ensures that 
$(u_{\theta,q}^\text{ad},v_{\theta,q}^\text{ad})$ are orthogonal to all 
eigenvectors of $\cal L$ with $E_n\ne 0$. 

Perturbation of the soliton can be expanded in the complete basis vectors
\bea
\left(\begin{array}{c} 
\delta\phi
\\
\delta\phi^*
\end{array}\right)&=&
\frac{\theta'-\theta}{i}
\left(\begin{array}{c} 
u_\theta 
\\
v_\theta
\end{array}\right)+P_\theta
\left(\begin{array}{c} 
u_\theta^{\rm ad} 
\\
v_\theta^{\rm ad} 
\end{array}\right)
\cr
&&
+\frac{q'-q}{i}
\left(\begin{array}{c} 
u_q 
\\
v_q
\end{array}\right)+P_q
\left(\begin{array}{c} 
u_q^{\rm ad} 
\\
v_q^{\rm ad} 
\end{array}\right)
\cr
&&
+\sum_{n,E_n>0}\left[b_n\left(\begin{array}{c} 
u_n 
\\
v_n
\end{array}\right)
+b_n^*\left(\begin{array}{c} 
v_n^* 
\\
u_n^*
\end{array}\right)
\right],
\label{pertbexp}
\eea
where real $q'$ and $\theta'$ describe translation of the soliton and shift of
its global phase, respectively, 
while $P_q$ and $P_\theta$ (also real) are momentum of
the CM of the soliton and momentum conjugate to the global phase, respectively.
The momentum $P_\theta=N'-N$ represents deviation from the average total number
of particles $N$.
Deformation of the soliton shape is described by complex variables $b_n$.
Substituting \eqref{pertbexp} into \eqref{tbe} and projecting on 
the basis vectors results in a set of equations
\bea
\partial_t \theta'&=&\frac{P_\theta}{M_\theta}+2
\la \partial_N\phi_0|V\phi_0\ra,\label{1} \\
\partial_t P_\theta&=&0,\label{2}  \\
\partial_t q'&=&\frac{P_q}{M_q},\label{3}  \\ 
\partial_t P_q&=&-\int dz\;|\phi_0(z-q)|^2\;\partial_zV(z),\label{4}  \\
i\partial_t b_n&=&E_n\;b_n+s_n,\label{5} 
\eea
where real-valued
\be
s_n=\la u_n|S\ra+\la v_n|S^*\ra.
\label{sn}
\ee
Equation~\eqref{1} describes linear evolution of the global phase and it is
possible to obtain $\theta'(t)=\theta={\rm const}$ by a proper choice of 
$P_\theta$. The latter is a constant of motion, see \eqref{2}.
We consider a weak disorder potential when $\sigma_0<\hl$. Therefore the force acting on
the CM, which is the force acting on a single particle convoluted with the soliton
profile \eqref{4}, is small and it oscillates around zero as a function of $q$. 
Thus, Eqs.~\eqref{3}-\eqref{4} imply that, if we choose $P_q(0)=0$ and such a
$q$ that $\int dz |\phi_0(z-q)|^2\partial_zV=0$, then $q'(t)=q={\rm const}$.

\subsection{Deformation of the soliton shape}
\label{meanb}

We have seen that in a disorder potential 
the CM of the soliton can be fixed and its global phase 
can be constant. Let us now concentrate on the set of Eqs.~\eqref{5} which 
describe changes in the soliton shape due to the presence of a disorder 
potential. Solving Eqs.~\eqref{5} with an assumption that initially the bright 
soliton is unperturbed, i.e. $b_n(0)=0$, we obtain 
\bea
\delta\phi&=&\sum_{n,E_n>0}\frac{s_n}{E_n}\left[
\left(e^{-iE_n t}-1\right)\;u_n(z-q)\right.
\cr
&&\left.+\left(e^{iE_n t}-1\right)\;v_n^*(z-q)
\right].
\eea
The lowest energy of the Bogoliubov modes in the case of the bright
soliton is $E_1=|\mu|=N^2/8$ \cite{ueda}.  Thus a large gap in energy separates the soliton from the 
Bogoliubov modes. These modes are delocalized and
describe radiation of the soliton. The energy spectrum can be well approximated 
by a shifted free particle dispersion relation
\be
E_n\approx \frac{2\pi^2}{L^2}n^2+|\mu|,
\ee
where $n$ is integer and $L$ stands for the size of a box in which we consider
our system. Moreover, due to the radiation character of the modes
\bea
|u_n+v^*_n|&\le& \frac{1}{\sqrt{L}}, \\
|s_n|&\le& |V_0|\sqrt{\frac{N\hl}{2L}}.
\eea
The latter inequality is obtained taking a rectangular  profile  of size $\hl$
for the bright soliton. Finally, with
$\sin^2(E_nt/2)\le1$ and $\sum_n 1/E_n\approx \int dn/E_n$, for 
deformation of the soliton shape,
\be
|\phi_0+\delta\phi|^2\approx |\phi_0|^2+ \phi_0\;\delta\phi^*+\phi_0^*\;\delta\phi,
\ee
we obtain the following estimate
\be
|\phi_0\;\delta\phi^*+\phi_0^*\;\delta\phi|\le 4|V_0|,
\ee
and if it is much smaller than $|\phi_0|^2\le 2|\mu|$, the shape of the soliton 
is negligibly changed. Hence, if we want the shape of the bright soliton 
to be unaffected by the presence of a disorder potential a sufficient condition 
is 
\be
|V_0|\ll|\mu|.
\ee

Note that the upper bound on  $V_0$ requires the potential to be sufficiently smooth, in particular the case of a $\delta$-correlated 
disorder potential is excluded by this condition \cite{kivshar}.

\subsection{Dziarmaga approach}
\label{meanc}

In Sec.~\ref{meana}, equations of motion for a bright soliton in
the presence of a weak disorder potential have been obtained using
 the perturbative expansion \eqref{pertbexp}.
Consequently the long time evolution of the CM of the
soliton for $P_q(0)\ne 0$ cannot be described by these equations. Indeed, 
after a finite time $|q'(t)-q|>\hl$ and the perturbative approach breaks down. 
Similar problem may occur in the case of the $\theta'$ variable. 

We will be interested in a quantum description of the bright soliton where states
corresponding to superposition of the CM position over a distance much larger
than $\hl$ will be considered. Therefore we need a method that allows us to
describe non-perturbative displacement of the soliton. To this end we adopt 
Dziarmaga approach introduced in a problem of quantum diffusion of a dark 
soliton \cite{dziarmaga04}. Following Ref.~\cite{dziarmaga04} we do not perform
a linear expansion of a perturbed soliton wave-function around fixed $q$ and
$\theta$ like in \eqref{pertbexp} but we treat $q$ and $\theta$ themselves as
dynamical variables

\bea
\left(\begin{array}{c} 
\phi
\\
\phi^*
\end{array}\right)&=&
\left(\begin{array}{c} 
\phi_0
\\
\phi_0^*
\end{array}\right)+P_\theta
\left(\begin{array}{c} 
u_\theta^{\rm ad}
\\
v_\theta^{\rm ad}
\end{array}\right)
+P_q
\left(\begin{array}{c} 
u_q^{\rm ad} 
\\
v_q^{\rm ad} 
\end{array}\right)
\cr
&&
+\sum_{n,E_n>0}\left[b_n\left(\begin{array}{c} 
u_n 
\\
v_n
\end{array}\right)
+b_n^*\left(\begin{array}{c} 
v_n^* 
\\
u_n^*
\end{array}\right)
\right].
\label{exp}
\eea
Note that now if $q(t)$ and $\theta(t)$ are changing in time all modes also 
evolve because they depend on $q$ and $\theta$, e.g. $u^{\rm ad}_\theta=u^{\rm
ad}_\theta(z-q(t))$. Substituting \eqref{exp} into energy functional
\eqref{enfunc} supplemented with the $\int dz V|\phi|^2$ term and keeping terms
of order ${\cal O}(P^2,b^2,PV,bV)$ only, we obtain the effective Hamiltonian
\bea
H&=&\frac{P_q^2}{2M_q}+\int dz\;V(z)\;|\phi_0(z-q)|^2 \cr
&& +\frac{P_\theta^2}{2M_\theta}+2P_\theta\la \partial_N\phi_0|V\phi_0\ra \cr
&& +\sum_{n,E_n>0}\left(E_nb_n^*b_n+(b_n+b_n^*)s_n\right),
\label{clham}
\eea
which generates the following equations motion 
\bea
\partial_t \theta&=&\frac{\partial H}{\partial P_\theta}=\frac{P_\theta}{M_\theta}+2
\la \partial_N\phi_0|V\phi_0\ra,  \label{n1}\\
\partial_t P_\theta&=&-\frac{\partial H}{\partial \theta}=0,  \\
\partial_t q&=&\frac{\partial H}{\partial P_q}=\frac{P_q}{M_q},  \\ 
\partial_t P_q&=&-\frac{\partial H}{\partial q}\approx-\int dz\;|\phi_0(z-q)|^2\;
\partial_zV(z), \label{n4}\\
i\partial_t b_n&=&\frac{\partial H}{\partial b_n^*}=E_n\;b_n+s_n.  \label{n5}
\eea
In \eqref{n4} we have neglected terms 
$P_\theta\partial_q\la \partial_N\phi_0|V\phi_0\ra$ and 
$(b_n+b_n^*)\partial_q s_n$ because they are of order of ${\cal O}(PV,bV)$ while
in the equations we keep the linear terms only. Strictly speaking in order to show
that the pairs of variables in \eqref{n1}-\eqref{n5} 
are canonically conjugate one should switch to
Lagrangian formalism of the problem, however, as the result is obvious, we have
skipped it, see \cite{dziarmaga04}.

Equations~\eqref{n1}-\eqref{n5} possess a form identical to  
\eqref{1}-\eqref{5}. However, $q$ and $\theta$ 
present in $\phi_0$ and $s_n$ on the right hand side of the current 
equations are not fixed and evolve in time. It introduces couplings between $q$
and $\theta$ and $b_n$ degrees of freedom which were absent in 
\eqref{1}-\eqref{5}.
Inserting solutions of \eqref{n1}-\eqref{n5} into \eqref{exp} 
we can obtain long distance propagation of a bright soliton including 
possible changes of its shape, something not possible with the expansion
\eqref{pertbexp}. 

The Hamiltonian \eqref{clham} cannot be used for extremely
large momentum of the CM. That is, it is valid provided 
$P_q\hl/N\ll 1$, compare \eqref{exp} and \eqref{adj}. For the case of large
$P_q$ see \cite{kivshar}.
Note also, that due to the fact $M_\theta$ is negative, see \eqref{masses},
the bright soliton \eqref{bs} is a saddle point of the energy functional 
(\ref{enfunc}). It has, however, no consequences since $P_\theta=N'-N$ is 
a constant of motion. 

\section{Quantum description}
\label{qdes}

From the point of view of quantum mechanics 
the classical ground state solution \eqref{bs} breaks $U(1)$ gauge and
translation symmetries of the quantum many Hamiltonian
\cite{dziarmaga04}. That is, the quantum
Hamiltonian commutes with $\hat U=e^{i\hat N\theta}$ and, 
in the absence of a disorder potential, also with the translation operator.
In the Bogoliubov description; the $\theta$ and $q$ 
degrees of freedom appear as zero energy modes and, 
thanks to the Dziarmaga approach, we know how to properly describe
arbitrarily large changes in $\theta$ and $q$.  

The quantum mechanical version of \eqref{clham} reads
\bea
\hat H&=&\frac{\hat P_q^2}{2M_q}+\int dz\;V(z)\;|\phi_0(z-\hat q)|^2 \cr
&& +\frac{\hat P_\theta^2}{2M_\theta}+2\hat P_\theta\la \partial_N\phi_0|V\phi_0\ra \cr
&& +\sum_{n,E_n>0}\left(E_n\hat b_n^\dagger\hat b_n+(\hat b_n+\hat b_n^\dagger)
s_n\right),
\label{qmham}
\eea
where
\bea
\hat P_q&=&-i\partial_q, \\
\hat P_\theta&=&\hat N-N=-i\partial_\theta,
\eea
and
\bea
\left[\hat q,\hat P_q\right]&=& i, \\
\left[\hat \theta,\hat P_\theta\right]&=&i, \\
\left[\hat b_n,\hat b_m^\dagger\right]&=&\delta_{nm}.
\eea

Because $[\hat P_\theta,\hat H]=0$ we can choose a state $|N\ra$ 
of the many body system with exactly $N$ particles 
where $\hat P_\theta|N\ra=0$. If we consider the Bogoliubov vacuum state of
the quasi-particle operators, i.e. $\hat b_n|0_b\ra=0$, such a state will 
be very weakly coupled to other eigenstates of the 
$\sum_n E_n\hat b_n^\dagger\hat b_n$ operator because
the coupling strengths $s_n$ are, for a weak disorder potential, 
much smaller than the large energy gap for quasi-particle excitations
$E_1=|\mu|=N^2/8$ \cite{ueda}. 
Hence, the effective Hamiltonian that describes the CM motion
reduces to
\bea
\hat H_q&=&\la N;0_b|\hat H|N;0_b\ra
\cr
&=&
\frac{\hat P_q^2}{2N}+\int dz\;V(z)\;|\phi_0(z-\hat q)|^2,
\label{hamq}
\eea
where we have inserted explicit expression for $M_q$.
In the following we will use the Hamiltonian \eqref{hamq} in analyzing of
Anderson localization of the CM of a bright soliton.
Second order contributions, with respect to the coupling
to the quasiparticle modes, to the effective Hamiltonian (43) are of
the order of $NV_0^2/\mu$ and they can be neglected for the parameters
of the system used in the present paper.

\section{Anderson localization}
\label{al}

We discuss in more detail the Anderson localization in the so called optical speckle potential,
as realized e.g. in the experiment~\cite{Billy2008}. The potential originates
from the light shifts experienced by the the atoms in the laser light detuned from the resonance.
In effect, the potential $V(z)\propto\alpha|E(z)|^2$ is proportional 
to the intensity of the local field $E(z)$ and to the atomic polarizability 
$\alpha$, whose sign depends on the detuning of the external light frequency 
from the atomic resonance.   

\begin{figure}
\centering
\includegraphics*[width=0.9\linewidth]{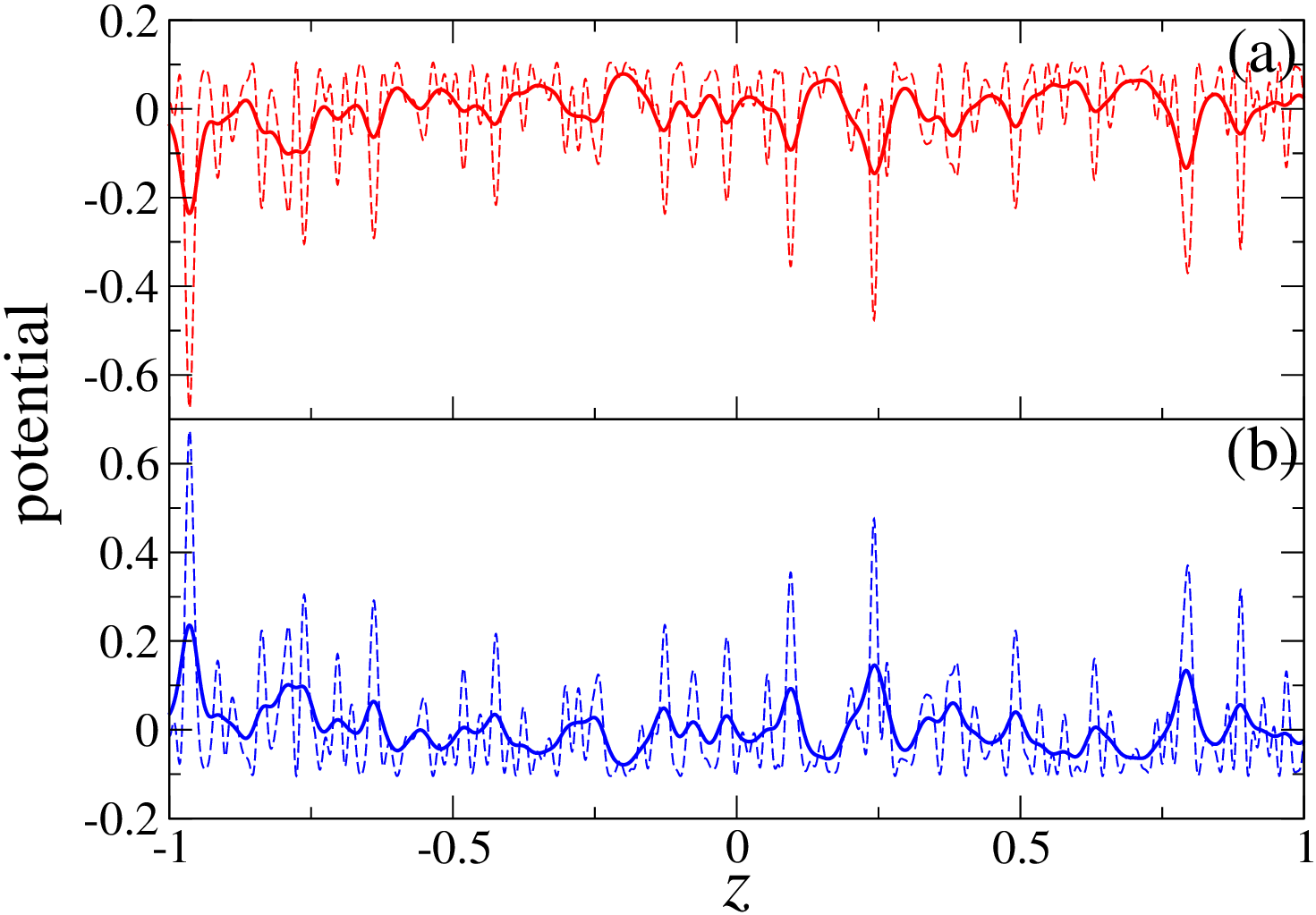}
\caption{Dashed lines: bare potential $V(z)$, 
solid lines: convoluted potential, i.e.
$\int dz' V(z')|\phi_0(z'-z)|^2/N$. Panel (a) for the bare potential
amplitude $V_0=+0.1$ (red detuned laser case), 
panel (b) for $V_0=-0.1$ (blue detuned laser case). 
The correlation length of the bare potential is $\sigma_0=0.28\hl$ 
where $\hl=0.02$.
} 
\label{one}
\end{figure}

Any disordered potential is completely characterized by its correlation functions 
$\overline{V(z_1)\dots V(z_n)}$ where the overbar denotes an ensemble average over disorder realizations.  
The average potential value shifts the origin of energy and can always be set to zero, $\overline{V(z)}=0$. 
The pair correlator can be written as $\overline{V(z')V(z'+z)}=V_0^2 C(z/\sigma_0)$, 
where $V_0$ measures the potential strength, and $\sigma_0$ the spatial correlation length. For
a gaussian random process, higher order correlation functions are simple functions of the average
and the pair correlator. This is no longer the case for 
non-gaussian potentials that require to specify also higher-order correlations. 

\begin{figure}
\centering
\includegraphics*[width=0.9\linewidth]{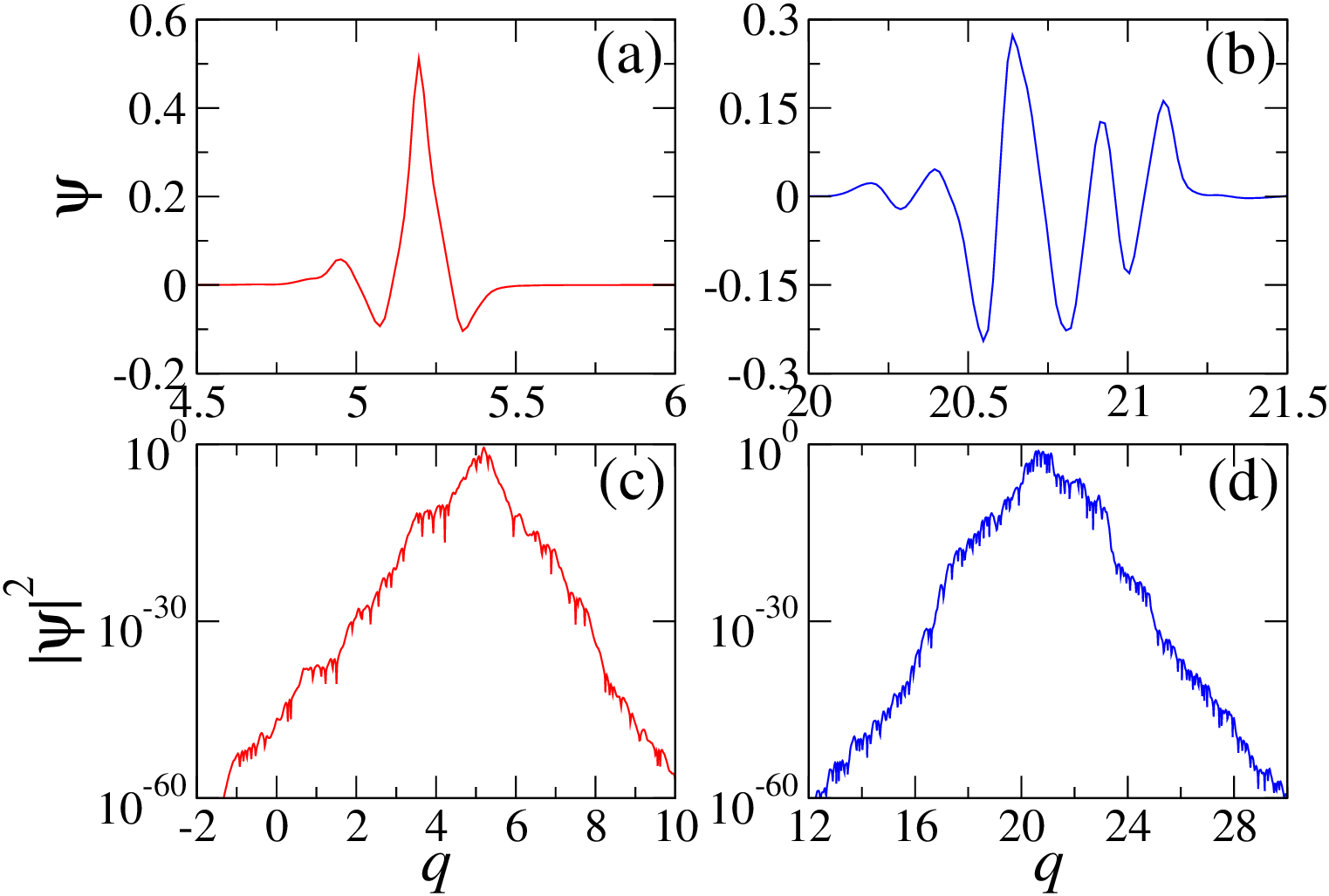}
\caption{Panels (a) and (b): eigenstates of the CM of a bright soliton; 
panels (c) and
(d): corresponding probability density in log scale. 
The eigenstates correspond
to the CM momentum $P_q\approx10$. The red detuned laser case is shown in (a)
and (c) while the blue detuned one in (b) and (d). The inverse localization
length is $\gamma=23\pm3$ (red detuned case) and 
$\gamma=16.5\pm0.6 $ (blue detuned case).
The parameters of the potentials are the same as in Fig.~\ref{one}.
} 
\label{two}
\end{figure}

An optical speckle potential is a good example of such a non-gaussian behaviour.    
At fixed detuning, the potential features either random peaks (the ``blue-detuned'' case) 
or wells (``red-detuned''). Even after shifting to $\overline{V(z)}=0$, 
the potential distribution 
is asymmetric (compare Fig.~\ref{one}), and the importance of odd moments can be probed experimentally by comparing 
the blue- and red-detuned cases for fixed amplitude $|V_0|$. 
The latter is determined by the laser 
strength, and we will use $|V_0|=8\cdot10^{-5}|\mu|=0.1$ in the following.   
The bare speckle potential has the pair correlation function $C(y)=[\sin (y)/y]^2$, 
with a correlation length that can be as short as 
$0.28\,\mu$m \cite{Billy2008} or $\sigma_0=0.0056$ in our units. We shall use this
value in the following. The CM of the soliton feels, however, not the bare potential, but rather 
its convolution with the soliton shape, see Eq.~\eqref{hamq}. The convoluted effective potential
$\int dz' V(z')|\phi_0(z'-z)|^2/N$ (the $N$ factor in the denominator is due to the
normalization of $\phi_0$) is also shown in Fig.~\ref{one}. While the convolution makes the
potential smoother it is apparent that it remains quite asymmetric, thus we may expect
that the non-gaussian character (in particular non-vanishing odd moments) shows up in 
the properties of the system. For that reason we compare the results for both red- and blue-
detuned potential of similar amplitude.

\begin{figure}
\centering
\includegraphics*[width=0.9\linewidth]{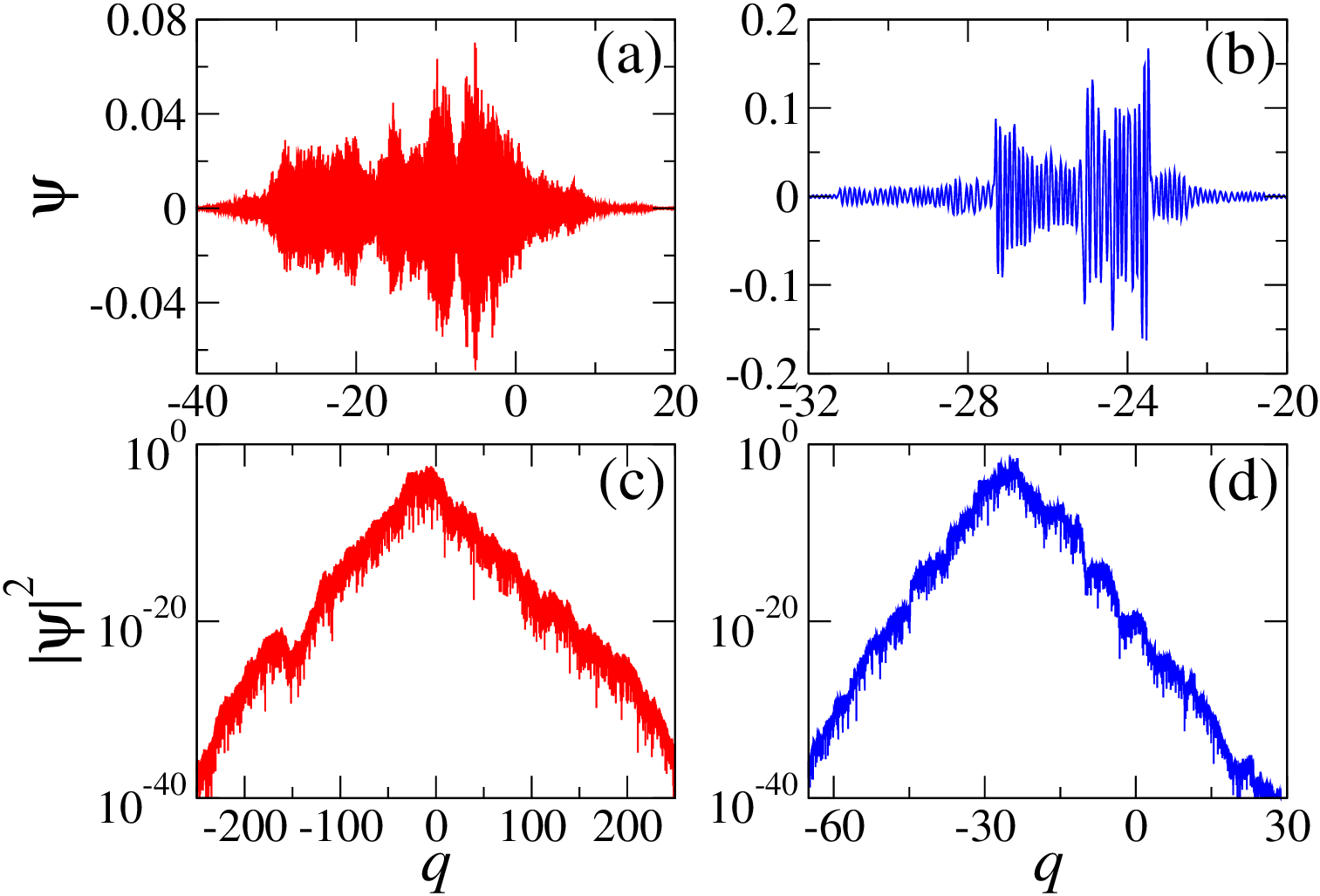}
\caption{Panels (a) and (b): eigenstates of the CM of soliton; panels (c) and
(d): corresponding probability density in log scale. The eigenstates correspond
to the CM momentum $P_q\approx50$. The red detuned laser case is shown in (a)
and (c) while the blue detuned one in (b) and (d). The inverse localization
length is $\gamma=0.27\pm0.03 $ (red detuned case) and 
$\gamma= 1.8\pm0.1$ (blue detuned case).
The parameters of the potentials are the same than in Fig.~\ref{one}.
} 
\label{three}
\end{figure}

The generic properties of  Anderson localization in 1D \cite{vantiggelen1999} allow us to
 expect that all the eigenstates of \eqref{hamq} are exponentially localized, i.e., have a
 typical shape with the overall envelope
 \be
 |\Psi|^2\propto \exp\left[-\gamma(P_q)|q-q_0|\right],
 \ee  
with $q_0$ being the mean position while $\gamma(P_q)$ is naturally referred to as the inverse
localization length. It depends on the eigenenergy of the state $E$, or writing $P_q\approx\sqrt{2NE}$
on the associated momentum $P_q$. By diagonalizing the Hamiltonian \eqref{hamq} on a grid,
we obtain the wavefunctions that are represented in Fig.~\ref{two} and Fig.~\ref{three} for
two significantly different energies (momenta). Fig.~\ref{two} shows the probability densities for the 
CM of the soliton at relatively low energies, observe that the exponential envelope behaviour
is visible over several decades. Due to the tridiagonal form of the diagonalized matrix on the grid,
the errors are well under control and the accuracy seems not to be limited by  double precision arithmetics.
Observe that the inverse localization lengths obtained for red-detuned case and the blue-detuned
situation differ significantly, stronger localization is observed for the former case.

The situation is quite different at higher energies as shown in 
Fig.~\ref{three}. Observe that now
blue-detuned potential leads to a much stronger localization. Of course the inverse localization lengths
at high energies are much smaller than those depicted in Fig.~\ref{two}, in fact, at sufficiently
high energies $\gamma(P_q)$ decays exponentially with $P_q$ as observed by us before \cite{ours}.

\begin{figure}
\centering
\includegraphics*[width=0.9\linewidth]{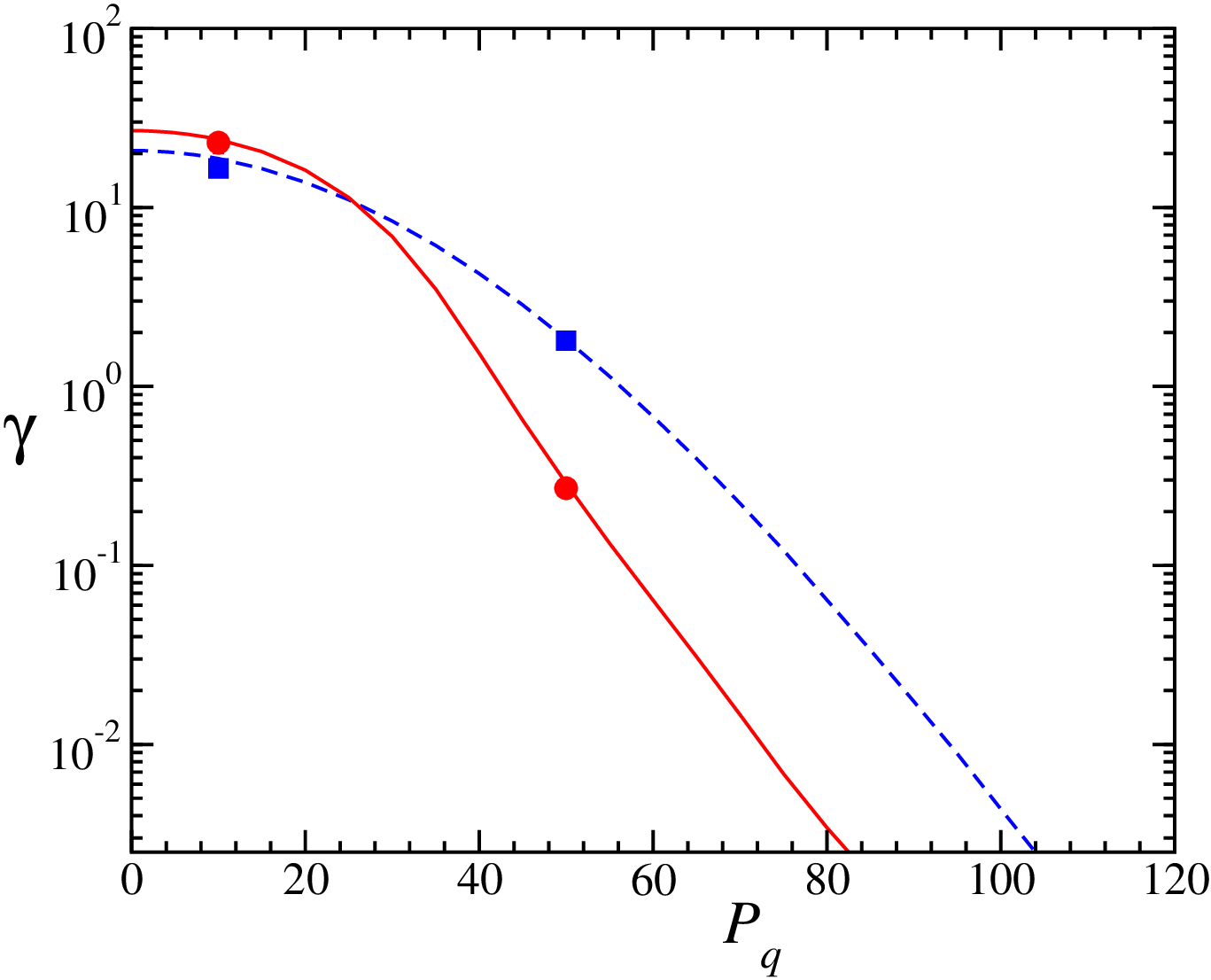}
\caption{Inverse localization length as a function of the momentum for red-detuned (red-line, solid)
and blue-detuned (blue-line, dashed) potentials obtained using the transfer matrix technique. 
The red dots as well as the blue
squares correspond to the wavefunctions shown in Fig.~\ref{two} and Fig.~\ref{three}. Observe
the exponential decay of $\gamma$ for sufficiently large $P_q$.
} 
\label{four}
\end{figure}

The inverse localization lengths shown as lines in Fig.~\ref{four} are obtained by a transfer matrix technique
\cite{MacKinnon1981} and quite nicely agree with values obtained from exact diagonalization. Clearly there
is a striking difference between the two cases of red-detuned and blue-detuned potentials exemplifying
its non-gaussian character and the importance of higher moments, 
in particular third moments. This in turn indicates
that the application of the celebrated Born approximation 
\cite{vantiggelen1999,ours} which considers the two lowest moments only is deemed to fail in our case
despite the fact that the potential is very weak, smooth and thus, at first glance one could naively
expect the Born approximation to perform quite well.

With exponentially localized eigenstates, one can now consider the dynamics, e.g., the spread of an initially
localized wavepacket. As shown by us elsewhere \cite{ours}, one can expect an algebraic localization of
such a CM wavepacket. For realistic parameters, 
localization occurs on a timescale of seconds making
the experimental verification of the localization feasible. We refer the reader to \cite{ours} for
details.

\section{Conclusions}

Using the Bogoliubov expansion and treating the zero modes non-perturbatively, we have shown in detail how to
obtain the effective quantum Hamiltonian which governs the motion of the center of mass of the bright soliton
in a weak and smooth potential without affecting the soliton shape. When this potential is of the disorder
type one may expect to observe Anderson localization of the CM motion. The optical speckle potential
was considered as a realistic example. It turns out that localization properties of wavefunctions strongly
depend on the sign of the potential (red- or blue- detuning). 
This indicates that, even for a weak potential,
applicability of the Born approximation is limited 
and the quantitative predictions depend on higher 
correlation functions of the disorder potential. 
Anderson localization of the CM of a bright soliton
should be experimentally observable.

\section*{Acknowledgements}

We are grateful for the privilege of delightful lively discussions with Cord M\"uller.  
Support within Polish Government scientific funds (for years 2008-2011 -- KS and 
2009-2012 -- JZ)
as a research project and by Marie Curie ToK project COCOS 
(MTKD-CT-2004-517186) is acknowledged. The research has been conducted
within LFPPI network.



\end{document}